\pageno=1                                      
\input mauche.mac                              
%
%
%
\font\ninerm=cmr9
\font\eightrm=cmr8
\font\sixrm=cmr6
\font\ninei=cmmi9
\font\eighti=cmmi8
\font\sixi=cmmi6
\skewchar\ninei='177 \skewchar\eighti='177 \skewchar\sixi='177
\font\ninesy=cmsy9
\font\eightsy=cmsy8
\font\sixsy=cmsy6
\skewchar\ninesy='60 \skewchar\eightsy='60 \skewchar\sixsy='60

\font\ninebf=cmbx9
\font\eightbf=cmbx8
\font\sixbf=cmbx6
\font\ninett=cmtt9
\font\eighttt=cmtt8
\hyphenchar\tentt=-1 
\hyphenchar\ninett=-1
\hyphenchar\eighttt=-1
\font\ninesl=cmsl9
\font\eightsl=cmsl8
\font\nineit=cmti9
\font\eightit=cmti8
\newskip\ttglue
\def\tenpoint{\def\rm{\fam0\tenrm}%
  \textfont0=\tenrm \scriptfont0=\sevenrm \scriptscriptfont0=\fiverm
  \textfont1=\teni \scriptfont1=\seveni \scriptscriptfont1=\fivei
  \textfont2=\tensy \scriptfont2=\sevensy \scriptscriptfont2=\fivesy
  \textfont3=\tenex \scriptfont3=\tenex \scriptscriptfont3=\tenex
  \def\it{\fam\itfam\tenit}%
  \textfont\itfam=\tenit
  \def\sl{\fam\slfam\tensl}%
  \textfont\slfam=\tensl
  \def\bf{\fam\bffam\tenbf}%
  \textfont\bffam=\tenbf \scriptfont\bffam=\sevenbf
   \scriptscriptfont\bffam=\fivebf
  \def\tt{\fam\ttfam\tentt}%
  \textfont\ttfam=\tentt
  \tt \ttglue=.5em plus.25em minus.15em
  \normalbaselineskip=12pt
  \let\sc=\eightrm
  \let\big=\tenbig
  \setbox\strutbox=\hbox{\vrule height8.5pt depth3.5pt width0pt}%
  \normalbaselines\rm}
\def\ninepoint{\def\rm{\fam0\ninerm}%
  \textfont0=\ninerm \scriptfont0=\sixrm \scriptscriptfont0=\fiverm
  \textfont1=\ninei \scriptfont1=\sixi \scriptscriptfont1=\fivei
  \textfont2=\ninesy \scriptfont2=\sixsy \scriptscriptfont2=\fivesy
  \textfont3=\tenex \scriptfont3=\tenex \scriptscriptfont3=\tenex
  \def\it{\fam\itfam\nineit}%
  \textfont\itfam=\nineit
  \def\sl{\fam\slfam\ninesl}%
  \textfont\slfam=\ninesl
  \def\bf{\fam\bffam\ninebf}%
  \textfont\bffam=\ninebf \scriptfont\bffam=\sixbf
   \scriptscriptfont\bffam=\fivebf
  \def\tt{\fam\ttfam\ninett}%
  \textfont\ttfam=\ninett
  \tt \ttglue=.5em plus.25em minus.15em
  \normalbaselineskip=10pt 
  \let\sc=\sevenrm
  \let\big=\ninebig
  \setbox\strutbox=\hbox{\vrule height8pt depth3pt width0pt}%
  \normalbaselines\rm}
\def\eightpoint{\def\rm{\fam0\eightrm}%
  \textfont0=\eightrm \scriptfont0=\sixrm \scriptscriptfont0=\fiverm
  \textfont1=\eighti \scriptfont1=\sixi \scriptscriptfont1=\fivei
  \textfont2=\eightsy \scriptfont2=\sixsy \scriptscriptfont2=\fivesy
  \textfont3=\tenex \scriptfont3=\tenex \scriptscriptfont3=\tenex
  \def\it{\fam\itfam\eightit}%
  \textfont\itfam=\eightit
  \def\sl{\fam\slfam\eightsl}%
  \textfont\slfam=\eightsl
  \def\bf{\fam\bffam\eightbf}%
  \textfont\bffam=\eightbf \scriptfont\bffam=\sixbf
   \scriptscriptfont\bffam=\fivebf
  \def\tt{\fam\ttfam\eighttt}%
  \textfont\ttfam=\eighttt
  \tt \ttglue=.5em plus.25em minus.15em
  \normalbaselineskip=9pt
  \let\sc=\sixrm
  \let\big=\eightbig
  \setbox\strutbox=\hbox{\vrule height7pt depth2pt width0pt}%
  \normalbaselines\rm}
%
\def\headtype{\ninepoint}                 
\def\abstracttype{\ninepoint}             
\def\captiontype{\ninepoint}              
\def\footnotetype{\ninepoint}             
\def\reftype{\ninepoint}                  
\def\refit{\it}                           
\font\chaptitle=cmr10 at 11pt             
\rm                                       

%
%
\parindent=0.25in                         
\parskip=0pt                              
\baselineskip=12pt                        
\hsize=4.25truein                         
\vsize=7.445truein                        
\hoffset=1in                              
\voffset=1in                           

\newskip\sectionskipamount                
\newskip\aftermainskipamount              
\newskip\subsecskipamount                 
\newskip\firstpageskipamount              
\newskip\capskipamount                    
\newskip\ackskipamount                    
\sectionskipamount=0.2in plus 0.09in
\aftermainskipamount=6pt plus 6pt         
\subsecskipamount=0.1in plus 0.04in
\firstpageskipamount=3pc
\capskipamount=0.1in
\ackskipamount=0.15in
\def\sectionskip{\vskip\sectionskipamount}
\def\aftermainskip{\vskip\aftermainskipamount}
\def\subsecskip{\vskip\subsecskipamount} 
\def\firstpageskip{\vskip\firstpageskipamount}
\def\capskip{\hskip\capskipamount}

%
%
\nopagenumbers                            
\newcount\firstpageno                     
\firstpageno=\pageno                      
\newcount\chapno                          

\def\rightheadline{\headtype\phantom{\folio}\hfil\runningtitletext\hfil\folio}
\def\leftheadline{\headtype\folio\hfil\runningauthortext\hfil\phantom{\folio}}
\headline={\ifnum\pageno=\firstpageno\hfil
           \else
              \ifdim\ht\topins=\vsize           
                 \ifdim\dp\topins=1sp \hfil     
                 \else
                     \ifodd\pageno\rightheadline\else\leftheadline\fi
                 \fi
              \else
                 \ifodd\pageno\rightheadline\else\leftheadline\fi
              \fi
           \fi}

\def\bottomnumber{\hss\tenrm[\folio]\hss}
\footline={\ifnum\pageno=\firstpageno\bottomnumber\else\hfil\fi}

%
%
%
%
\outer\def\mainsection#1
    {\vskip 0pt plus\smallskipamount\sectionskip
     \message{#1}\vbox{\noindent{\bf#1}}\nobreak\aftermainskip\noindent}
 
\outer\def\subsection#1
    {\vskip 0pt plus\smallskipamount\subsecskip
     \message{#1}\vbox{\noindent{\bf#1}}\nobreak\smallskip\nobreak\noindent}
 

\def\title#1{{\chaptitle\leftline{#1}}}
\def\name#1{\leftline{#1}}
\def\affiliation#1{\leftline{\it #1}}
\def\abstract#1{{\abstracttype \noindent #1 \smallskip\vskip .1in}}
\def\ref{\noindent \parshape2 0truein 4.25truein 0.25truein 4truein}
\def\caption{\noindent \captiontype
             \parshape=2 0truein 4.25truein .125truein 4.125truein}

\def\footnote#1{\edef\fspafac{\spacefactor\the\spacefactor}#1\fspafac
      \insert\footins\bgroup\footnotetype
      \interlinepenalty100 \let\par=\endgraf
        \leftskip=0pt \rightskip=0pt
        \splittopskip=10pt plus 1pt minus 1pt \floatingpenalty=20000
        \textindent{#1}\bgroup\strut\aftergroup\strut\egroup\let\next}
\skip\footins=12pt plus 2pt minus 4pt 
\dimen\footins=30pc 

%
%

\def\@{\spacefactor 1000}

\def\,{\pcomma} 
\def\pcomma{\relax\ifmmode\mskip\thinmuskip\else\thinspace\fi}

\def\oversim#1#2{\lower0.5ex\vbox{\baselineskip=0pt\lineskip=0.2ex
     \ialign{$\mathsurround=0pt #1\hfil##\hfil$\crcr#2\crcr\sim\crcr}}}


\def\runningtitletext{THE WINDS OF CATACLYSMIC VARIABLES}
\def\runningauthortext{C.~W.~MAUCHE and J.~C.~RAYMOND}

\null
\firstpageskip

\vbox to 0pt{\vskip -2.7cm
\hbox to \hsize{1997, in {\it Cosmic Winds and the Heliosphere\/}, ed.~J.~R.~Jokipii,\hfil }
\hbox to \hsize{C.~P.~Sonett, \& M.~S.~Giampapa (Tucson: Univ.~of Arizona Press)     \hfil }
\vss}
{\baselineskip=14pt
\title{THE WINDS OF CATACLYSMIC VARIABLES}
}

\vskip .3truein
\name{CHRISTOPHER W.~MAUCHE}
\affiliation{Lawrence Livermore National Laboratory}
\vskip .1truein
\leftline{and}
\vskip .1truein
\name{JOHN C.~RAYMOND}
\affiliation{Harvard-Smithsonian Center for Astrophysics}
\vskip .3truein

\abstract{%
We present an observational and theoretical review of the winds of cataclysmic
variables (CVs). Specifically, we consider the related problems of the
geometry, ionization state, and mass-loss rate of the winds of CVs. We present
evidence of wind variability and discuss the results of studies of eclipsing
CVs. Finally, we consider the properties of accretion disk wind and magnetic
wind models. Some of these models predict substantial angular momentum loss,
which could affect both disk structure and binary evolution.}


\mainsection{{I}.~~Introduction}

Cataclysmic variables (CVs) are a large, diverse class of short-period 
semi-detached mass-exchanging binaries composed of a white dwarf, a late-type
(G, K, or M) main-sequence secondary, and, typically, an accretion disk.
The various CV subtypes include novae, which are powered by the thermonuclear
burning of material on the surface of the white dwarf, and dwarf novae and
nova-like variables, which are powered by the accretion of material
transferred from the secondary to the white dwarf. For recent reviews of CVs,
refer to Cropper (1990), Livio (1994), and C\'ordova (1995).

Among the nova-like variables, there exists a class in which the magnetic
field of the white dwarf is strong enough to influence the flow of material
accreted by the white dwarf. These magnetic nova-like variables are
subdivided into polars (AM~Her stars), whose white dwarfs have spin periods
which are synchronous with the orbital period, and intermediate polars
(DQ~Her stars), whose spin periods are less than the orbital period. In
polars, the strength of the magnetic field of the white dwarf, the small size
(short period) of the binary, and the degree of synchronization do not allow
an accretion disk to form. In intermediate polars, the lower strength of the
magnetic field of the white dwarf, the larger size (longer period) of the
binary, and the lack of spin-orbit synchronization allow an accretion
disk to form. This disk maintains an essentially Keplerian velocity profile
at large radii, but is subsequently disrupted as it moves within the
magnetosphere of the white dwarf.

Dwarf novae form a distinctive CV subtype characterized by the frequency,
duration, and magnitude of their outbursts. The outbursts of dwarf novae have
recurrence times of tens to hundreds of days, durations of between 1 and 10
days, and increases of between 3 and 5 mag in their visual brightness. These
properties are thought to result from an instability either in the rate of
mass exchange from the secondary to the accretion disk of the white dwarf, or
in the rate of mass transfer through the accretion disk itself (Smak 1984;
Cannizzo 1993). Nonmagnetic nova-like variables appear to be best described
as dwarf novae ``stuck'' in outburst.

There are two sources of winds in CVs. The first, restricted to nonmagnetic
nova-like variables and dwarf novae in outburst, is evidenced by the \pcygni
\ profiles of their ultraviolet resonance lines. In analogy with the winds
of early-type stars, the winds of these CVs are generally thought to be driven
by radiation pressure, a process which results in high-velocity outflows with
mass-loss rates limited by the conservation of momentum to $\Mdot _{\rm wind}
\lax L/V_{\infty }c$, where $L$ is the luminosity of the radiation field and
$V_{\infty }$ is the terminal velocity of the wind. In CVs, these quantities
can be written as $L = \zeta \, G\Mwd\Mdot/\Rwd $ and $V_{\infty } = f\,
V_{\rm esc} = f\, (2G\Mwd/\Rwd)^{1/2}$, respectively, where $\Mwd $ and
$\Rwd $ are the mass and radius of the white dwarf, respectively, and
$\zeta $ and $f$ are constants of order unity. Hence, the mass-loss rate of
the wind is $\Mdot _{\rm wind} \lax \zeta f^{-1}\, (G\Mwd/2c^2\Rwd )^{1/2}\,
\Mdot \sim 0.01 \, \zeta f^{-1}\, \Mdot$. The second source of winds in CVs is
the solar-type wind of the secondary. Although this wind has no {\it direct\/}
observational manifestation, it is of great theoretical interest, as it is
appealed to to drive the secular evolution of the binary. For earlier reviews
of the winds of CVs, refer to C\'ordova and Howarth (1987) and Drew and Kley
(1993).

\mainsection{{I}{I}.~~Geometry and Mass-Loss Rate}

The study of the winds of CVs began with the launch of \IUE , the first UV
spectroscopic satellite with sufficient sensitivity and spectral resolution
to first discover (Heap \etal \ 1978), and subsequently to study, the \pcygni
\ profiles of the ultraviolet resonance lines of CVs: namely \nv \ $\lambda
1240$, \siiv \ $\lambda 1400$, and \civ \ $\lambda 1550$. The ``early'' papers
reporting \IUE \ observations of ``high $\Mdot $'' CVs---classical novae,
nova-like variables, and dwarf novae in outburst (e.g., Krautter \etal \
1981; Klare \etal \ 1982; Greenstein and Oke 1982; Guinan and Sion 1982;
C\'ordova and Mason 1982, 1985; Sion 1985)---all contain a discussion of
the \pcygni \ profiles of these lines. 

Observationally, the \pcygni \ profiles of CVs differ from the profiles of
early-type stars in two significant ways, both of which are illustrated by
the spectra shown in Fig.~1. First, whereas the deepest part of the
blueshifted absorption component of the line profiles of early-type stars is
near the terminal velocity, the deepest part of the line profiles of CVs is
near zero velocity. Second, the absorption components of the line profiles
of early-type stars are often are black, whereas in CVs they are at most 
roughly half the depth of the continuum. As judged by the velocity of the blue
edge of the absorption component of the profiles, the terminal velocity of the
wind is roughly $5000~\kms $. This coincidence between the terminal velocity
of the wind and the escape velocity of the white dwarf suggests that the winds
of CVs originate from deep in the accretion disk, the boundary layer between
the disk and the surface of the white dwarf, or the white dwarf itself
(C\'ordova and Mason 1982).

As in early-type stars, the \pcygni \ profiles of the ultraviolet resonance
lines of CVs are formed by scattering in their winds (Drew and Verbunt 1985).
Specifically, the line profiles are formed by the scattering of continuum
photons of frequency $\nu $ which come into resonance with the
radially-accelerating wind at a point in velocity space given by
$V(r) = (\nu - \nu _0)c/\nu _0$, where $\nu _0$ is the rest frequency of the
transition. The material in the wind moving toward the observer in front of the
continuum source scatters photons out of the line of sight, and so produces a
blueshifted absorption feature. From all other regions of the wind (except the
region hidden behind the continuum source), we receive photons which have been
scattered into our line of sight. Since the projected velocities of these parts
of the wind  range from positive to negative, the emission feature so produced
is roughly symmetric about $\nu _0$. In the Sobolev approximation to the
radiation transfer in the expanding wind, the fraction of scattered photons
varies as $1-\exp[-\tau (r)]$, where the optical depth  $\tau (r) \propto
\Mdot_{\rm wind}\, \xi (r)\, r^{-2}\,  V(r)^{-1}\, [dV(r)/dr] ^{-1}$, where 
$V(r)$ and  $\xi (r)$ are respectively the velocity and ionization laws
characterizing the wind.

Possibly different wind geometries aside, the winds of CVs differ from the
winds of early-type stars in the source function of the scattered continuum,
which is the photosphere of the star in the former case, and the accretion
disk in CVs. Because of this differing source of illumination, and because of
the inclination dependence introduced by the presence of the accretion disk,
the theoretical \pcygni \ profiles of early-type stars (e.g., Olson 1978;
Castor and Lamers 1979) cannot be applied to the profiles of CVs. In order
to understand why this is so, it is instructive to consider the case of a
spherical wind emanating from the center of a luminous accretion disk. If
the disk is observed edge-on, no part of the wind lies in front of the disk,
and so no absorption results; the \pcygni \ profile consists simply of an
emission component whose full width is given by the terminal velocity of the
wind: $\Delta\nu = 2\nu _0V_{\infty }/c$. As the inclination from which the
disk is observed decreases, the fraction of the wind seen in projection
against the disk increases, thereby increasing the strength of the absorption
component of the line profile. At the same time, the strength of the emission
component decreases due to both the trade-off of emission for absorption
regions, and the corresponding increase of the volume of the wind hidden
behind the disk. In the limit of a face-on disk, roughly half the wind is
hidden behind the disk and roughly half is seen in projection against the
disk; the resulting profile has a strong absorption component and little or
no emission. This inclination dependence of the \pcygni \ profiles of CVs is
illustrated in Fig.~2.

Given this important difference between the \pcygni \ profiles of CVs and
early-type stars, and the possibility that the geometry of the winds of CVs
is more nearly bipolar than spherical, theoretical profiles appropriate to
the peculiar conditions of CVs must be constructed in order that well-founded
conclusions can be drawn concerning the nature of their winds. In addition to
the mass-loss rate, velocity law, and ionization law characterizing the wind,
one must specify the source function and geometry of the wind, as well as 
the source function of the scattered continuum, which is typically taken to
be given by the Shakura and Sunyaev (1973) temperature distribution of the
accretion disk---itself dependent on the mass-accretion rate and the mass
and radius of the white dwarf---and the emissivity of the Planck function:
$I_\lambda (r) = B_\lambda [T(r;\> \Mwd ,\, \Rwd ,\, \Mdot)]$. Given
this plethora of variables, the first models of the \pcygni \ profiles of CVs
(Drew 1987; Mauche and Raymond 1987) were designed with aerodynamic
simplicity: the wind was assumed to originate in the boundary layer or inner
disk, the ionization state was assumed to be constant with radius, and the
velocity law was assumed to be given by either a power law or linear function
of the radius.

In the context of these models, Drew (1987) and Mauche and Raymond (1987) came
to the following conclusions regarding the nature of the winds of CVs. First,
the wind can originate near the white dwarf only if its acceleration is very
slow compared to the winds of early-type stars. Specifically, the wind 
must not reach a significant fraction of its terminal velocity before
reaching a distance from the white dwarf comparable to the distance to the
peak of the distribution of the resonant continuum in the accretion disk,
which for  typical parameters is at $\sim$10 white dwarf radii. Second, the
mass-loss rate of the wind is $\sim {\rm few} \times 10^{-11}\, \xi_{\rm
C\thinspace IV}^{-1}\, \Msun\, \peryr $, where $\xi_{\rm C\thinspace IV}$ is
the ionization fraction of \civ . With typical mass-accretion rates of $\Mdot
\sim 10^{-8}\, \Msun\, \peryr $, the ionization fraction of \civ \ must then
be no less than a few tenths if the wind is driven by radiation pressure.
Finally, some type of bipolarity in either the wind density profile [e.g.,
$n(r,\theta ) = n(r,0)|\cos ^l\theta |$] or the radiation field [e.g., limb
darkening: $I(\theta ) = {1\over 2}(1+{3\over 2}|\cos\theta |)\, I(0)$] is
required to avoid the formation of an absorption component in the line
profiles of high-inclination systems.

\mainsection{{I}{I}{I}.~~Ionization State}

In order to fully characterize the winds of CVs, we also need to know their
ionization state. It is, after all, the absolute concentrations of the ions
which ultimately determine the wind's mass-loss rate, the relative
concentrations of the ions which determine the relative strengths of the
line profiles, and radial variations in the concentrations of the ions which,
in collusion with the velocity law, determine the shapes of the profiles.
Unfortunately, we have few observational handles on the ionization state of
the wind other than those provided by the \nv , \siiv , and \civ \ resonance
lines. \HUT \ has recently provided information for transitions in the
1200--912\AAA \ bandpass, but for only a few systems (e.g., Long \etal \
1991, 1994). Into this near vacuum, theory guides us.

Despite many theoretical attempts to understand the ionization state of the
winds of CVs, investigators repeatedly find that if the temperature and
luminosity of the radiation generated in the boundary layer is as given by
simple theory (e.g., Pringle 1977; Patterson and Raymond 1985), and if the
wind is smooth and represents a mass-loss rate which is a modest fraction of
the accretion rate, the wind will be ionized beyond the observed ionization
stages of \nv , \siiv , and \civ . Drew and Verbunt (1985) found that a
relatively slow velocity law, as well as variations in the temperature and
luminosity of the boundary layer, help to increase the ionization fractions
of the observed ions, but not by an amount sufficient to explain the
strengths of the line profiles. Kallman and Jensen (1985) suggested that the
resolution of the related problems of the ionization state of the wind and
the apparent discrepancies between observations and theoretical predictions
of the luminosity and temperature of the boundary layer radiation was to be
found in a wind of sufficient density to photoelectrically absorb the EUV and
soft \xray \ flux emitted by the boundary layer. However, they found that
the required mass-loss rate of such a wind exceeds the mass-accretion rate,
a possibility excluded by the conservation of energy, which limits the
mass-loss rate to $\Mdot _{\rm wind} \lax \zeta f^{-1}\, \Mdot$. A recent
\ROSAT \ observation of the dwarf nova SU~UMa in outburst shows an increased
absorbing column of about $10^{20}~{\rm cm}^{-2}$ compared with quiescence, a
column which does not drastically reduce the observable soft \xray s (Silber,
Vrtilek, and Raymond 1994). In another attempt to solve this problem, Mauche and
Raymond (1987) investigated the consequences of introducing shock compression
in the wind in an attempt to reduce its ionization state relative to that of
a smooth wind. In order to produce reasonably large \civ \ ionization
fractions, they found that mass-loss rates of $\Mdot _{\rm wind} \sim 0.3\,
\Mdot$ are required if shocks are present with velocities of $\sim 100~\kms
$. Although this solution allows a wind on energetic grounds, Mauche and
Raymond found that such a wind produces much larger UV line fluxes than are
observed and (in their view) is far too efficient in absorbing the soft \xray
s generated in the boundary layer.

Since that time, observations (Mauche \etal\ 1991; van Teeseling, Verbunt,
and Heise 1993; Mauche, Raymond, and Mattei 1995) and theory (Hoare and Drew
1991) have (again) called into question the parameters characterizing the
boundary layer radiation. Hoare and Drew (1993) have therefore considered wind
models with cool, luminous boundary layers and no boundary layer at all. They
found that the observed strength of the \civ \ and \nv \ lines can be matched by
models with luminous, cool ($kT=5$--9~\ev ) boundary layers or with disk-only
models if the accretion rate is $\Mdot \sim 2$--$4\times 10^{-8}\, \Msun\,
\peryr $. The softer radiation fields of these models allow low mass-loss rate
($\Mdot _{\rm wind} \sim 6\times 10^{-10}\, \Msun\, \peryr \sim 0.02\, \Mdot$)
winds, capable of being driven by radiation pressure. The presence of blueshifted
absorption in \siiv \ remains a problem, however, as does the presence of a
strong O~VI line (Long \etal \ 1991, 1994). A further difficulty is the origin
of the soft \xray s observed in SS~Cyg, U~Gem, VW~Hyi, and OY~Car (C\'ordova
\etal \ 1980, 1984; van der Woerd, Heise, and Bateson 1986; Naylor \etal \
1988). If the wind is driven by radiation pressure, a natural alternative for
the origin of the soft \xray s is shocks in the wind (Mauche and Raymond 1987;
Naylor \etal \ 1988). However, this solution fails to explain the production
of the quasi-periodic oscillations observed in the soft \xray \ flux of SS~Cyg,
U~Gem, and VW~Hyi (C\'ordova \etal \ 1980, 1984; van der Woerd \etal\ 1987,
Mauche 1995), as well as why the \xray s generated in the winds of CVs should
be so much cooler than the \xray s produced by the winds of O stars (e.g.,
Chlebowski, Harnden, and Sciortino 1989).

Clearly, more work is needed in this area. Much concerning the ionization state
of the wind could be done with the additional information from the far-UV
waveband supplied by \HUT \ and the EUV waveband supplied by \EUVE . Indeed,
\EUVE \ observations of the dwarf nova SS~Cyg in outburst (Mauche, Raymond, and
Mattei 1995) have produced results so different from theoretical expectations
that they may radically change our understanding of the boundary layers and
winds of CVs. First, whereas all models of boundary layer emission (e.g.,
Pringle 1977; Popham and Narayan 1995) predict dramatic changes in the effective
temperature with accretion rate, and whereas the brightness of the source rose
by a factor of over 100 during the \EUVE \ observations, the {\it shape\/} of
the EUV spectrum was constant. Second, whereas theory predicts that the
luminosities of the boundary and accretion disk should be comparable unless
the white dwarf is rotating near breakup, $L_{\rm bl}/L_{\rm disk}\sim 1$,
observations imply that this ratio is less than 0.1. Third, contrary to simple
theory, the EUV spectrum of SS~Cyg is a complex mixture of emission and
absorption features dominated by transitions of 5--7 times ionized Ne, Mg, and
Si. Work to understand these results is ongoing.

\mainsection{{I}{V}.~~Variability}

As is evident from the variability of the \pcygni \ profiles, the winds of
CVs are not time-steady. At one extreme, the winds are observed to develop
and subsequently to decay as the outbursts of dwarf novae evolve (e.g.,
Verbunt \etal \ 1984; la Dous \etal \ 1985; Hassall, Pringle, and Verbunt 1985;
Verbunt \etal \ 1987). At the other extreme, the winds are observed to vary
on timescales of hours down to tens of minutes (e.g., Prinja, Drew, and Rosen
1992); about as quickly as UV spectra can be obtained with \IUE . Variations
on the orbital phase are also observed; the most spectacular being the orbital
phase-dependent profile and equivalent width variations observed in the line
profiles of the dwarf nova YZ~Cnc (Drew and Verbunt 1988; see Fig.~3). The
series of observations precipitated by this discovery (e.g., Woods, Drew, and
Verbunt 1990; Woods \etal \ 1992) prove, if nothing else, that line profile
variations are common on the orbital period. The cause of these variations is
less than clear. In general, profile variations can be produced by an azimuthal
asymmetry in either the brightness distribution of the accretion disk, or in
the wind's geometry, velocity field, or ionization structure. The trick is
understanding how these asymmetries are produced, and why they are stable. The
answer may tell us something important about the accretion disk or the wind,
but little quantitative information exists at the present time to narrow the
selection of possibilities. 

One potentially important, but as yet little-studied, diagnostic observable of
the winds of CVs is the radial velocity variations of their \pcygni \ profiles.
Mauche (1991) discovered such variations in high-resolution \IUE \ spectra of
the nova-like variable IX~Vel. The radial velocity variations of the \pcygni \
profiles of the \siiv\ doublet of this system are well fit by $V_{\rm Si\,
IV}(\phi ) = -350 - 200\, \sin [2\pi(\phi - 0.1)]~\kms $. The difference
between this solution and the radial velocity solution of the white dwarf and
accretion disk is $V_{\rm wind}(\phi) \approx  -400 - 130\,  \sin [2\pi(\phi
- 0.75)]$: the radial velocity of the wind lags the white dwarf and accretion
disk by $90\degrees $. One interpretation of this result is that the velocity
of the wind in which the \siiv \ ion dominates varies according to  $V_{\rm
wind}(\varphi ) \approx 400 - 130\, \cos (2\pi\varphi )$, where $\varphi $ is
the azimuthal angle between the white dwarf and the secondary: the effective
velocity of the wind is lowest on the side facing the secondary and highest on
the side facing away. This type of variation may not be that uncommon:  Drew,
Hoare, and Woods (1991) observed velocity shifts in the \nv \ and \civ \
profiles of the dwarf nova DX~And; Mauche \etal \ (1994) found similar radial
velocity variations in the \nv , \siiv , \civ , and \heii \ emission lines
(presumably wind-dominated features) of the eclipsing nova-like variable V347
Pup (see Fig.~4). Given the diagnostic potential of these radial velocity
variations, it seems clear that much could be learned from radial velocity
studies of other CVs, particularly those for which the orbital ephemeris is
well known.

\mainsection{{V}.~~Eclipse Studies}

Unlike T~Tauri stars or AGN, CVs present the opportunity of eclipse studies
of the disk and wind. \IUE \ studies of UX~UMa, RW~Tri, OY~Car, Z~Cha, and
V347 Pup (Holm, Panek, and Schiffer 1982; King \etal\ 1983; C\'ordova and Mason
1985; Drew and Verbunt 1985; Naylor \etal \ 1988; Harlaftis \etal \ 1992{\it
a\/}, {\it b\/}; Mauche \etal \ 1994) all demonstrate that the ultraviolet
emission lines of CVs are not completely extinguished in eclipse, strongly
indicating that they are formed in a region with dimensions comparable in size
to the size of the secondary. Modulo the possible effects of superposed
absorption (Drew 1987), the relative depths of the eclipses of these lines
also give some indication of the relative dimensions of the regions in which
their profiles are formed. V347 Pup is representative: there, the  ratio of the
eclipse to mean out-of-eclipse flux of the \nv , \siiv , and \civ \ emission
lines are 35\%, 33\%, and 53\%, respectively (Mauche \etal \ 1994; see Fig.~4).
The weak eclipse of the \civ \ profile relative to the those of \nv \ and \siiv
\ indicates that the region of the wind in which the \civ \ ion dominates is
more extensive than for these other two ions; a conclusion which is consistent
with the stronger emission component of the \pcygni \ profile of \civ \ relative
to those of \nv \ and \siiv \ in lower inclination systems. It is not yet clear
whether carbon exists in the form of \civ \ throughout a larger fraction of the
outer wind than \nv \ or \siiv , or whether a wind with constant ionization
fractions would simply give $\tau \sim 1$ at a larger radius for the more
abundant \civ \ ion. The former scenario is more consistent with theoretical
expectations, since nitrogen and silicon will exist as N\thinspace IV and
Si\thinspace V in those regions of the wind where \civ \ dominates (see, e.g.,
Mauche and Raymond 1987).

In much the same way that eclipse mapping probes the brightness temperature
distribution of the accretion disks of CVs (see, e.g., Horne 1993), eclipse
studies also constrain the surface brightness distribution of the winds of
CVs, and hence constrain the wind's source function and geometry. Although
\IUE \ does not supply the combination of effective area and spectral and
temporal resolution required to do true eclipse mapping, some conclusions can
be drawn from the existing data. The case of V347 Pup is instructive: as
shown in Figs.~4 and 5({\it a\/}), the \civ \ emission profile decreases both
in strength and in width as the wind is eclipsed by the secondary. The same
effect is observed in the dwarf nova OY~Car (Naylor \etal \ 1988). In contrast,
as shown in Fig.~5({\it b\/}), the profiles of existing wind models (Drew
1987; Mauche \etal \ 1994) {\it in\/}crease in width in eclipse. Although this
difference between observations and models can be reconciled a number of ways,
the results for the optical emission profiles of V347 Pup suggest that the UV
emission profiles may be broadened significantly by rotation. If this is the
case, it implies that the wind arises from the surface of the accretion disk
rather than from near the white dwarf (Shlosman, Vitello, and Mauche 1995).

\mainsection{{V}{I}.~~Disk Winds}

Accretion disk winds actually have a number of qualities to recommend them
over a wind originating near the white dwarf. First, the gravitational
potential at the surface of the accretion disk is much lower than at the
surface of the white dwarf. This difference implies a reduction in the energy
required to drive the wind of a given mass-loss rate to escape velocity.
Second, since a disk wind arises from the same region responsible for the
resonant photons which produce the \pcygni \ profiles, lower mass-loss rates
are required to produce a given column density, and hence a given line
strength, than the mass loss-rate required of a wind originating near the
white dwarf. Third, since a disk wind arises from a spatially extended
region, the volume of the wind in which photons are resonantly scattered is
likely to be larger than the corresponding volume of a wind originating near
the white dwarf. The size of this volume has implications for the depth and
width of the eclipse of the line-forming region in systems in which the
secondary eclipses the white dwarf. Since the eclipse depths are observed to
be fairly small, an extended scattering region is very desirable. Fourth,
the $r^{-2}V(r)^{-1}$ increase in the density of the wind associated with a
spherically symmetric geometry is softened in a disk wind. This difference
could remove the high-density component of the wind overlying the boundary
layer and allow soft \xray s to emerge through the wind. 

Shlosman and Vitello (1993) and Vitello and Shlosman (1993) have recently
explored the properties of disk winds. They assume that the wind  originates
from an annulus of the accretion disk (see Fig.~6), that it rotates initially
with its local Keplerian value and conserves angular momentum as it
accelerates outward, and that its terminal velocity scales with the local
escape velocity. In total, nine free parameters (plus $\Mwd $, $\Rwd $,
$\Mdot $, and $\Mdot _{\rm wind}$) are required to fully characterize the
wind. The model calculates the wind ionization structure self-consistently
under the assumptions of constant temperature and local ionization equilibrium
and treats the radiation transfer in the Sobolev approximation.

Vitello and Shlosman find that disk winds are capable of providing good fits
to the line profiles of RW~Sex, RW~Tri, and V~Sge for reasonable parameters,
but note that {\it unique\/} fits are not possible, given the many degrees of
freedom in the model and the limited number of parameters characterizing the
profiles. They find that the introduction of rotation in the wind introduces
a radial shear in the velocity which decreases the optical depth, resulting
in a reduction in the line center intensity and a broadening of the emission
component of the line profile. In agreement with observations, significant
absorption does not occur at high inclination angles because the wind is
confined within an angle $\theta _{\rm max}$, which is typically taken to be
$65\degrees $. As expected, compared to winds which originate near the white
dwarf, disk winds require lower mass-loss rates to produce a given line
strength, and naturally explain the shallow line eclipse depths of eclipsing
CVs. Yet to be demonstrated is  whether a single manifestation of Shlosman and
Vitello's kinematic model can be applied successfully to a wide range of
systems with well-constrained white dwarf masses, accretion rates, and
inclination angles. As these authors point out, what is ultimately required is
a self-consistent dynamical model for the two-dimensional outflow from a disk.

\mainsection{{V}{I}{I}.~~Dynamics}

All of the models discussed above have assumed various wind geometries and
velocity laws in order to compute \pcygni \ profiles for diagnostic use.
In order to discuss dynamics, one first needs to specify the driving force.
The general similarity between the \pcygni \ profiles of CVs and early-type
stars has lulled most researchers into the belief that radiation pressure in
spectral lines drives the wind. However, the only attempt to model such a flow
resulted in velocities as high as those observed, but mass-loss rates which
were far too low, at least in comparison with spherical wind models (Raymond,
Van Ballegooijen, and Mauche 1988). The reason for the low predicted mass-loss
rate is that the critical point is determined by the behavior of the effective
gravity along the streamline, and it lies far above the disk where the density
is low. This problem {\it might\/} be overcome if a rapid change in ionization
structure modifies the driving force enough to lower the critical point to just
above the disk surface (see Vitello and Shlosman 1988).

The other dynamical models have employed magnetic fields. While Alfv\'en
waves (see MacGregor, this volume) or events similar to coronal mass
ejections (see Hundhausen, this volume) might well be important, the models
which have been developed so far rely on centripetal acceleration by a
large-scale magnetic field.  Blandford and Payne (1982) constructed a
self-similarity solution for application to AGN jets which has been applied
to CVs by Koen (1986) and Cannizzo and Pudritz (1988). In these models, the
field encounters the disk with an angle less than $60\degrees $ to the
equatorial plane, so that the rotating field accelerates material along the
field lines.  This acceleration continues out to the Alfv\'en radius,
$r_A$, where the field becomes too weak to enforce corotation.  Gas leaving
the disk at radius $r$ carries enough angular momentum to drive an accretion
rate $\Mdot = \Mdot_{\rm wind}\, r_A^2 / r^2$ through the disk, replacing
viscous transport of angular momentum through the disk as the fundamental
physical process responsible for accretion. The removal of mass and angular
momentum from the disk might account for the surprising flatness of the
far-UV spectra of some nova-like variables (la Dous 1991; Rutten, van Paradijs,
and Tinbergen 1992; Long \etal \ 1994). If the mass-loss rate of the wind is
large enough, it is possible that only a small fraction of the mass transferred
from the secondary reaches the white dwarf, providing at least in principle
an explanation for the faintness of the boundary layers in high-$\Mdot$ CVs.
A potential problem with the magnetically driven wind model is that the
large-scale magnetic field tends to diffuse outward through the disk, so that
some dynamo field generation is probably needed (Van Ballegooijen 1989). A
more complete model must also dispense with the  assumption that the field
lines are straight.  The field should be vertical where it passes through the
equatorial plane and curve outwards on a scale comparable to the disk
thickness.  It is the combination of the local field direction with the
density fall-off in the disk which actually determines the mass-loss rate
(Van Ballegooijen 1989).  Van Ballegooijen's model of the field geometry
also shows that the field lines are nearly vertical in the inner disk, so
that they are bent over enough to drive a wind only from the outer disk,
making it difficult to explain the \IUE \ and \HUT \ observations which
pertain more to the inner disk.

\mainsection{{V}{I}{I}{I}.~~Evolution}

Winds are crucial to many CVs because these binaries must lose angular
momentum to drive mass transfer.  Except for systems driven by nuclear
evolution (very long period systems), the companion would recede inside its
Roche lobe without angular momentum loss, and mass transfer would cease. 
Gravitational radiation carries away enough angular momentum to drive mass
transfer in the short period systems ($P \leq 2$~hr), but fails to
dispose of angular momentum fast enough to account for the moderate period
systems. In the generally accepted picture, magnetic braking by the wind of
the companion, a late-type star spinning synchronously with the orbital
period, drives the mass loss and the evolution toward shorter periods. No
direct measurements of this wind are available, but scaling from the
single-star magnetic braking rates derived from young clusters (Stauffer
1991) leads to good agreement with the observed mass-transfer rates (Verbunt
and Zwaan 1981; Rappaport, Verbunt and Joss 1983; Patterson 1984). In this
picture, magnetic braking ceases when the companion becomes fully convective
at a period of 3 hr and mass transfer switches off.  The system becomes
active again only after gravitational radiation brings the system back into
contact as the orbit shrinks and the period reaches 2 hr.  Some indirect
support for this picture comes from the observation of strong winds in dMe
stars (Mullan, this volume). A possible problem is the absence of the
expected drop in the \xray \ luminosity in field stars at M6, corresponding
to the companion in a 3 hr binary (Schmitt, this volume).  This makes the
sudden cessation of magnetic braking due to a drop in magnetic activity less
appealing.  In any case, the extrapolation of angular momentum loss rate,
$\Jdot $, from young stars to stars in 5 hr binaries is a plausible but very
risky undertaking (Charbonneau, this volume).

Magnetic braking by a disk wind is an alternative to magnetic braking by the
companion. The connection between the winds observed by \IUE \ and their
efficiency in disposing of angular momentum is rather tenuous, however. 
First, the magnetic field is unknown, and second, the observed winds arise
from the inner disk, while most of the angular momentum resides in the outer
disk. Cannizzo and Pudritz (1988) estimate the scaling of $\Jdot$ with period
based on simplified assumptions, and they find that braking by a disk wind
might well account for the observed mass-transfer rate in systems having
periods above 3 hr. In order to shut off mass transfer at a period of 3
hr, they assume that when the companion becomes fully convective its
magnetic activity ceases, and there is then no seed field to generate a disk
magnetic field.  An interesting aspect of this model is its potential
feedback.  A high mass-transfer rate causes the wind which carries away
angular momentum, which in turn shrinks the orbit and causes mass transfer.
Livio and Pringle (1994) consider the possibility that this mechanism causes
self-excited mass transfer. The period gap occurs when the companion becomes
completely covered by starspots, so that the mass transfer is inhibited long
enough for the companion to  shrink inside its Roche surface, shutting off
the feedback loop.  This model explains VY Scl ``anti-dwarf nova'' events as
a temporary cessation of mass transfer when a single starspot covers the
Lagrangian point.

A possible serious drawback of models which rely on magnetic braking by a disk
wind is the likelihood of instability. Lubow, Papaloizou, and Pringle (1994)
show that accretion through a disk driven by modest angular momentum loss in
a wind is unstable. There is also a possibility that if the specific angular
momentum of the disk wind is very large and the mass ratio of the system is
small, loss of angular momentum from the system could lead to dynamical mass
transfer (see, for instance, equation 1 of Melia and Lamb 1987). There is as
yet no direct evidence for the large-scale magnetic fields needed to transport
angular momentum, or for winds from the outer disk where most of the angular
momentum resides, but more detailed studies of CV \pcygni \ profiles, their
eclipses, and their orbital variations may define the wind geometry well enough
to discriminate among the competing models.

\mainsection{{I}{X}.~~The Future}

The era of \IUE \ studies of the winds of CVs is coming to a close. Although
it remains amazingly productive, \IUE \ is getting old, is in increasingly
poor health, and may be put down in the near future for lack of operating
funds. However, \IUE\ will not soon be forgotten; active research will continue
with the enormous archive that has accumulated over the past 17 plus years.
A small subset of this archive has appeared in the form of a 500-page
catalogue of low-resolution CV spectra (la Dous 1990), but the real power
of the archive lies in the ability to do {\it quantitative\/} analysis of
thousands of spectra of over one hundred CVs. A small number of
high-resolution \IUE \ spectra also reside in the archive (e.g., C\'ordova
1986; Mauche, Raymond, and C\'ordova 1988; Mauche 1991; Prinja and Rosen 1995),
but these spectra seem to have failed the promise they first appeared to offer
to reveal the secrets of the winds of CVs.

For additional observational progress to be made, higher quality (higher
signal-to-noise ratio, higher spectral resolution, higher temporal resolution)
spectra are required. When it can be wrestled away from the ``big boys''
working on $H_0$ and $q_0$, \HST \ provides the required properties in the
near-UV, but the importance of the far-UV and the EUV to the study of the
winds of CVs cannot be overstated. \HST \ has already observed the eclipsing
CVs UX~UMa (Mason \etal \ 1995) and OY~Car; \HUT \ has observed UX~UMa, Z~Cam
(Long  \etal \ 1991), and IX~Vel (Long \etal \ 1994); {\it ORFEUS\/} recently
observed Z~Cam and V3885~Sgr; \EUVE \ has observed SS~Cyg (Mauche, Raymond, and
Mattei 1995), U~Gem, and VW~Hyi. With some luck, these and subsequent data sets
will prove much of what we understand about the winds of CVs to be wrong.

\medskip

The authors are indebted to F.~C\'ordova, J.~Drew, T.~Kallman, and F.~Verbunt
for innumerable discussions relating to the winds of CVs. This work was
performed under the auspices of the U.S.~Department of Energy by Lawrence
Livermore National Laboratory under contract No.~W-7405-Eng-48 and under NASA
grant NAGW-528 to the Smithsonian Astrophysical Observatory.

\vfill\eject
\null


\vskip .50in
\centerline{\bf REFERENCES}
\vskip .25in
\reftype{
\ref{Blandford, R.~D., and Payne, D.~G. 1982. 
     Hydromagnetic flows from accretion discs and the production of radio jets.
     \mnras\ 199:883--903.}

\ref{Cannizzo, J.~K. 1993.
     The limit cycle instability in dwarf nova accretion disks.
     In {\refit Accretion Disks in Compact Stellar Systems\/},
     ed.\ J.\ C.\ Wheeler (Singapore: World Scientific), pp.~6--40.}

\ref{Cannizzo, J.~K., and Pudritz, R.~E. 1988. 
     A new angular momentum loss mechanism for cataclysmic variables.
     \apj\ 327:840--844.}

\ref{Castor, J.~I., and Lamers, H.~J.~G.~L.~M. 1979. 
     An atlas of theoretical P~Cygni profiles.
     \apjs\ 39:481--511.}

\ref{Chlebowski, T., Harnden, F.~R., and Sciortino, S. 1989.
     The Einstein X-ray observatory catalog of O-type stars.
     \apj\ 341:427--455.}

\ref{C\'ordova, F.~A. 1986.
     The promise of high resolution UV spectroscopy for understanding the winds
     of cataclysmic variables.
     In {\it The Physics of Accretion Onto Compact Objects\/},
     eds.\ K.\ O.\ Mason, M.\ G.\ Watson, and N.\ E.\ White
     (Berlin: Springer-Verlag), pp.~339--355.}

\ref{C\'ordova, F.~A. 1995.
     Cataclysmic variable stars.
     In {\refit X-ray Binaries\/},
     eds.\ W.\ H.\ G.\ Lewin, J.\ van Paradijs, and E.\ P.\ J.\ van den Heuvel
     (Cambridge: Cambridge Univ.~Press), pp.~331--388.}

\ref{C\'ordova, F.~A., Chester, T.~J., Mason, K.~O., Kahn, S.~M., and Garmire,
     G.~P. 1984. 
     Observations of quasi-coherent soft X-ray oscillations in U~Geminorum and
     SS~Cygni.
     \apj\ 278:739--753.}

\ref{C\'ordova, F.~A., Chester, T.~J., Tuohy, I.~R., and Garmire, G.~P. 1980. 
     Soft X-ray pulsations from SS~Cygni.
     \apj\ 235:163--176.}

\ref{C\'ordova, F.~A., and Howarth, I.~D. 1987. 
     Accretion onto compact stars in binary systems.
     In {\refit Exploring the Universe with the IUE Satellite\/},
     eds.\ Y.~Kondo, \etal \
     (Dordrecht: D.~Reidel), pp.~395--426.}

\ref{C\'ordova, F.~A., and Mason, K.~O. 1982. 
     High-velocity winds from a dwarf nova during outburst.
     \apj\ 260:716--721.}

\ref{C\'ordova, F.~A., and Mason, K.~O. 1985. 
     High-velocity winds in close binaries with accretion disks.
     II. - The view along the plane of the disk.
     \apj\ 290:671--682.}

\ref{Cropper, M. 1990. 
     The polars.
     \ssr\ 54:195--295.}

\ref{Drew, J.~E. 1987. 
     Inclination and orbital-phase-dependent resonance line-profile
     calculations applied to cataclysmic variable winds.
     \mnras\ 224:595--632.}

\ref{Drew, J.~E., Hoare, M.~G., and Woods, J.~A. 1991. 
     Ultraviolet observations of the long-period dwarf nova DX~And in outburst.
     \mnras\ 250:144--151.}

\ref{Drew, J.~E., and Kley, W. 1993.
     Mass loss and the boundary layer.
     In {\refit Accretion Disks in Compact Stellar Systems\/},
     ed.\ J.\ C.\ Wheeler (Singapore: World Scientific), pp.~212--242.}

\ref{Drew, J.~E., and Verbunt, F. 1985. 
     Investigation of a wind model for cataclysmic variable ultraviolet
     resonance line emission.
     \mnras\ 213:191--213.}

\ref{Drew, J.~E., and Verbunt, F. 1988. 
     Regular orbital variations in the ultraviolet resonance lines of YZ~Cnc.
     \mnras\ 234:341--351.}

\ref{Greenstein, J.~L., and Oke, J.~B. 1982. 
     RW~Sextantis, a disk with a hot, high-velocity wind.
     \apj\ 258:209--216.}

\ref{Guinan, E.~F., and Sion, E.~M. 1982. 
     Ultraviolet spectroscopy of the nova-like variable V3885 Sagittarii
     (=CD $-42^\circ 14462$).
     \apj\ 258:217--223.}

\ref{Harlaftis, E.~T., \etal \ 1992{\refit a\/}. 
     UV spectroscopy of Z~Chamaeleontis.
     I. - Time dependent dips in superoutburst.
     \mnras\ 257:607--619.}

\ref{Harlaftis, E.~T., \etal \ 1992{\refit b\/}. 
     UV spectroscopy of Z~Chamaeleontis.
     II. - The 1988 January normal outburst.
     \mnras\ 259:593--603.}

\ref{Hassall, B.~J.~M., Pringle, J.~E., and Verbunt, F. 1985. 
     Dwarf novae in outburst - Monitoring WX~Hydri with IUE.
     \mnras\ 216:353--363.}

\ref{Heap, S.~R., \etal \ 1978. 
     IUE observations of hot stars - HZ~43, BD $+75^\circ 325$, NGC~6826,
     SS~Cygni, Eta~Carinae.
     \nature\ 275:385--388.}

\ref{Hoare, M.~G., and Drew, J.~E. 1991. 
     Boundary-layer temperatures in high accretion rate cataclysmic variables.
     \mnras\ 249:452--459.}

\ref{Hoare, M.~G., and Drew, J.~E. 1993. 
     The ionization state of the winds from cataclysmic variables without
     classical boundary layers.
     \mnras\ 260:647--662.}

\ref{Holm, A.~V., Panek, R.~J., and Schiffer III, F.~H. 1982. 
     Ultraviolet spectrum variability of UX~Ursae Majoris.
     \apjl\ 252:L35--L37.}

\ref{Horne, K. 1993.
     Eclipse mapping of accretion disks.
     In {\refit Accretion Disks in Compact Stellar Systems\/},
     ed.\ J.\ C.\ Wheeler (Singapore: World Scientific), pp.~117--147.}

\ref{Kallman, T.~R., and Jensen, K.~A. 1985. 
     Soft X-rays, winds, and the cataclysmic variable boundary-layer problem.
     \apj\ 299:277--285.}

\ref{King, A.~R., Frank, J., Jameson, R.~F., and Sherrington, M.~R. 1983. 
     Phase-dependent UV spectra of UX~Ursae Majoris.
     \mnras\ 203:677--683.}

\ref{Klare, G., \etal \ 1982. 
     IUE observations of dwarf novae during active phases.
     \aap\ 113:76--84.}

\ref{Koen, C. 1986. 
     Angular momentum transport in the magnetospheres of cataclysmic variable
     accretion discs.
     \mnras\ 223:529--538.}

\ref{Krautter, J., \etal \ 1981. 
     IUE spectroscopy of cataclysmic variables.
     \aap\ 102:337--346.}

\ref{la Dous, C. 1990. 
     A catalogue of low-resolution IUE spectra of dwarf novae and nova-like
     stars.
     \ssr\ 52:203--706.}

\ref{la Dous, C. 1991. 
     New insights from a statistical analysis of IUE spectra of dwarf novae
     and nova-like stars. I. - Inclination effects in lines and continua.
     \aap\ 252:100--122.}

\ref{la Dous, C., \etal \ 1985. 
     Dwarf novae in outburst - Simultaneous ultraviolet and optical
     observations of RU~Pegasi and TZ~Persei.
     \mnras\ 212:231--243.}

\ref{Livio, M. 1994.
     Topics in the theory of cataclysmic variables and X-ray binaries.
     In {\refit Interacting Binaries\/}, eds.\ H.\ Nussbaumer and A.\ Orr
     (Berlin: Springer-Verlag), pp.~135--262.}

\ref{Livio, M., and Pringle, J.~E. 1994. 
     Star spots and the period gap in cataclysmic variables.
     \apj\ 427:956--960.}

\ref{Long, K.~S., \etal \ 1991. 
     Spectroscopy of Z Camelopardalis in outburst with the Hopkins Ultraviolet
     Telescope.
     \apjl\ 381:L25--L29.}

\ref{Long, K.~S., Wade, R.~A., Blair, W.~P., Davidson, A.~F., and Hubeny, I.
     1994. 
     Observations of the bright novalike variable IX~Velorum with the Hopkins
     Ultraviolet Telescope.
     \apj\ 426:704--715.}

\ref{Lubow, S.~H., Papaloizou, J.~C.~B., and Pringle, J.~E. 1994.
     On the stability of magnetic wind-driven accretion discs.
     \mnras\ 268:1010--1014.}

\ref{Mason, K.~O., \etal \ 1995.
     Eclipse observations of an accretion disc wind.
     \mnras\ 274:271--286.}

\ref{Mauche, C.~W. 1991. 
     High-resolution IUE spectra of the nova-like variable IX Velorum.
     \apj\ 373:624--632.}

\ref{Mauche, C.~W. 1995.
     EUVE photometry of SS~Cygni: dwarf nova outbursts and oscillations.
     In {\refit Proceedings of IAU Colloquium No.~152---Astrophysics in the
     Extreme Ultraviolet\/}, eds.~S.~Bowyer and B.~Haisch (Cambridge:
     Cambridge Univ.~Press), in press.

\ref{Mauche, C.~W., and Raymond, J.~C. 1987. 
     IUE observations of the dwarf nova HL~Canis Majoris and the winds of
     cataclysmic variables.
     \apj\ 323:690--713.}

\ref{Mauche, C.~W., Raymond, J.~C., and C\'ordova, F.~A. 1988. 
     Interstellar absorption lines in high-resolution IUE spectra of
     cataclysmic variables.
     \apj\ 335:829--843.}

\ref{Mauche, C.~W., Raymond, J.~C., and Mattei, J.~A. 1995.
     EUVE observations of the anomalous 1993 August outburst of SS~Cygni.
     \apj\ 446:842--851.}

\ref{Mauche, C.~W., Wade, R.~A., Polidan, R.~S., van der Woerd, H., and
     Paerels, F.~B.~S. 1991. 
     On the X-ray emitting boundary layer of the dwarf nova VW~Hydri.
     \apj\ 372:659--663.}

\ref{Mauche, C.~W., \etal \ 1994. 
     Optical, IUE, and ROSAT observations of the eclipsing nova-like variable
     V347 Puppis (LB~1800).
     \apj\ 424:347--369.}

\ref{Melia, F., and Lamb, D.~Q. 1987.
     Dynamical mass transfer in cataclysmic variables. 
     \apjl\ 321:L139--L143.}

\ref{Naylor, T., \etal \ 1988. 
     The 1985 May superoutburst of the dwarf nova OY~Carinae.
     II. - IUE and EXOSAT observations.
     \mnras\ 231:237--255.}

\ref{Olson, G.~L. 1978. 
     An analysis of ultraviolet resonance lines and the possible existence of
     coronae in O stars.
     \apj\ 226:124--137.}

\ref{Patterson, J. 1984. 
     The evolution of cataclysmic and low-mass X-ray binaries.
     \apjs\ 54:443--493.}

\ref{Patterson, J., and Raymond, J.~C. 1985. 
     X-ray emission from cataclysmic variables with accretion disks.
     II. - EUV/soft X-ray radiation.
     \apj\ 292:550--558.}

\ref{Popham, R., and Narayan, R. 1995.
     Accretion disk boundary layers in cataclysmic variables. I. Optically
     thick boundary layers.
			  \apj\ 442:337--357.}

\ref{Pringle, J.~E. 1977. 
     Soft X-ray emission from dwarf novae.
     \mnras\ 178:195--202.}

\ref{Prinja, R.~K., Drew, J.~E., and Rosen, S.~R. 1992. 
     Variability in the UV resonance lines of the cataclysmic variable
     V795 Herculis (PG 1711+336).
     \mnras\ 256:219--228.}

\ref{Prinja, R.~K., and Rosen, S.~R. 1995.
     High-resolution IUE spectroscopy of fast winds from cataclysmic variables.
     \mnras\ 273:461--474.

\ref{Rappaport, S., Joss, P.~C., and Verbunt, F. 1983.
     A new technique for calculations of binary stellar evolution, with
     application to magnetic braking.
     \apj\ 275:713--731.}

\ref{Raymond, J.~C., Van Ballegooijen, A.~A., and Mauche, C.~W. 1988.
     Structure and dynamics of cataclysmic variable winds.
     \baas\ 20:1020 (abstract).}

\ref{Rutten, R.~G.~M., van Paradijs, J., and Tinbergen, J. 1992. 
     Reconstruction of the accretion disk in six cataclysmic variable stars.
     \aap\ 260:213--226.}

\ref{Shakura, N.~I., and Sunyaev, R.~A. 1973. 
     Black holes in binary systems. Observational appearance.
     \aap\ 24:337--355.}

\ref{Shlosman, I., and Vitello, P. 1993. 
     Winds from accretion disks - Ultraviolet line formation in cataclysmic
     variables.
     \apj\ 409:372--386.}

\ref{Shlosman, I., Vitello, P., and Mauche, C.~W. 1995. 
     Rotating winds from accretion disks in cataclysmic variables: eclipse
     modeling of V347 Puppis.
     \apj\ submitted.}

\ref{Silber, A., Vrtilek, S.~D., and Raymond, J.~C. 1994. 
     Concurrent X-ray and optical observations of two dwarf novae during
     eruption.
     \apj\ 425:829--834.}

\ref{Sion, E.~M. 1985. 
     On the nature of the UX Ursa Majoris-type nova-like variables:
     CPD $-48^\circ 1577$.
     \apj\ 292:601--605.}

\ref{Smak, J. 1984.
     Outbursts of dwarf novae.
     \pasp\ 96:5--18.}

\ref{Stauffer, J.~R. 1991.
     Rotational velocities of low mass stars in young clusters.
     In {\refit Angular Momentum Evolution of Young Stars\/},
     eds.\ S.\ Catalano and J.\ R.\ Stauffer (Dordrecht: Kluwer),
     pp.~117--134.}

\ref{Van Ballegooijen, A.~A. 1989. 
     Magnetic fields in the accretion disks of cataclysmic variables.
     In {\refit Accretion Disks and Magnetic Fields in Astrophysics\/},
     ed.\ G.\ Belvedere (Dordrecht: Kluwer), pp.~99--106.}

\ref{van der Woerd, H., Heise, J., and Bateson, F. 1986. 
     The soft X-ray superoutburst of VW Hydri.
     \aap\ 156:252--260.}

\ref{van der Woerd, H., \etal \ 1987. 
     Discovery of soft X-ray oscillations in VW Hydri.
     \aap\ 182:219--228.}

\ref{van Teeseling, A., Verbunt, F., and Heise, J. 1993. 
     The nature of the X-ray spectrum of VW Hydri.
     \aap\ 270:159--164.}

\ref{Verbunt, F., Hassall, B.~J.~M., Pringle, J.~E., Warner, B., and Marang,
     F. 1987. 
     Multiwavelength monitoring of the dwarf nova VW~Hydri.
     III. - IUE observations.
     \mnras\ 225:113--130.}

\ref{Verbunt, F., \etal \ 1984. 
     Dwarf novae in outburst - Simultaneous ultraviolet and optical
     observations of UZ Serpentis, RX~Andromedae and AH Herculis.
     \mnras\ 210:197--221.}

\ref{Verbunt, F., and Zwaan, C. 1981. 
     Magnetic braking in low-mass X-ray binaries.
     \aap\ 100:L7--L9.}

\ref{Vitello, P., and Shlosman, I. 1988. 
     Line-driven winds from accretion disks.
     I. - Effects of the ionization structure.
     \apj\ 327:680--692.}

\ref{Vitello, P., and Shlosman, I. 1993. 
     Ultraviolet line diagnostics of accretion disk winds in cataclysmic
     variables.
     \apj\ 410:815--828.}

\ref{Woods, J.~A., Drew, J.~E., and Verbunt, F. 1990. 
     Time dependence of the UV resonance lines in the cataclysmic variables
     SU UMa, RX And and 0623+71.
     \mnras\ 245:323--330.}

\ref{Woods, J.~A., Verbunt, F., Collier Cameron, A., Drew, J.~E., and Piters,
     A. 1992. 
     Time dependence of the UV resonance lines in the cataclysmic variables
     YZ~Cnc, IR~Gem and V3885~Sgr.
     \mnras\ 255:237--242.}

}
\vfill\eject
\null


\vskip .50in
\centerline{\bf FIGURE CAPTIONS}
\vskip .25in

\caption{Figure~1.\capskip Typical short-wavelength \IUE \ spectra of the
         nova-like variables IX\thinspace Vel and V3885\thinspace Sgr.}

\caption{Figure~2.\capskip Theoretical \pcygni \ profiles for a specific
         mass-loss rate and velocity law for nine values of the
         inclination. From Mauche and Raymond (1987).}

\caption{Figure~3.\capskip \civ \ line profile variations as a function of
         orbital phase for the dwarf nova YZ~Cnc; flux density is measured
         in units of $10^{-12}$ erg cm$^{-2}$ s$^{-1}$ \AA $^{-1}$. Adapted
         from Drew and Verbunt (1988).}

\caption{Figure~4.\capskip \civ \ emission line flux, gaussian width, and
         radial velocity as a function of orbital phase for the
         nova-like variable V347 Pup. Best sinusoidal fit to radial
         velocity variation is overplotted; line flux is measured in
         units of $10^{-13}$ erg cm$^{-2}$ s$^{-1}$. From Mauche \etal
         \ (1994).}

\caption{Figure~5.\capskip ({\it a\/}) Eclipse ({\it lower curve\/}) and mean
         out-of-eclipse ({\it upper curve\/}) \civ \ line profiles of
         V347 Pup. ({\it b\/}) Model line profiles at orbital phases
         $\phi = 0$, 0.03, 0.04, 0.05, 0.06, 0.08, and 0.25. From
         Mauche \etal \ (1994).}

\caption{Figure~6.\capskip Geometry for a disk wind. From Shlosman and Vitello
         (1993).}

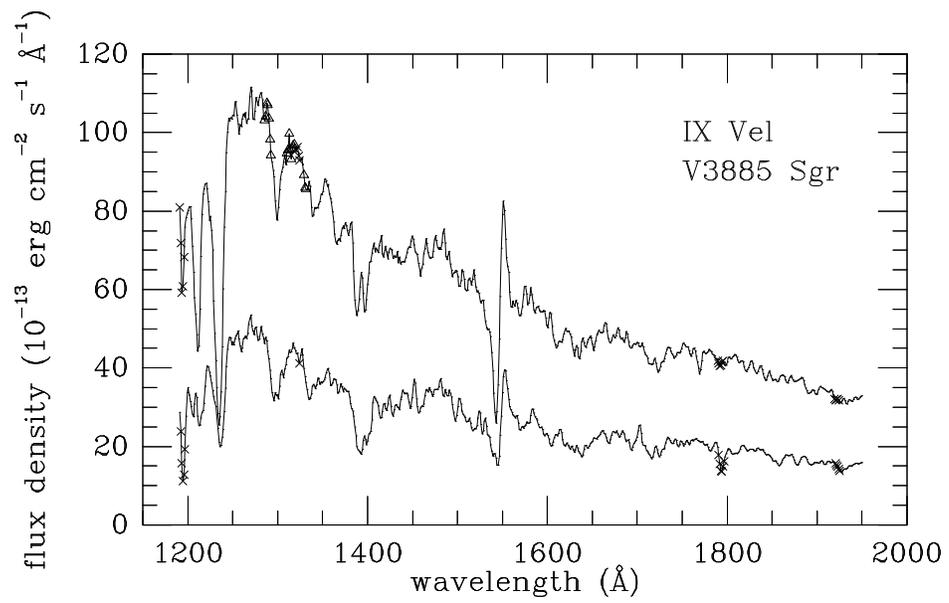
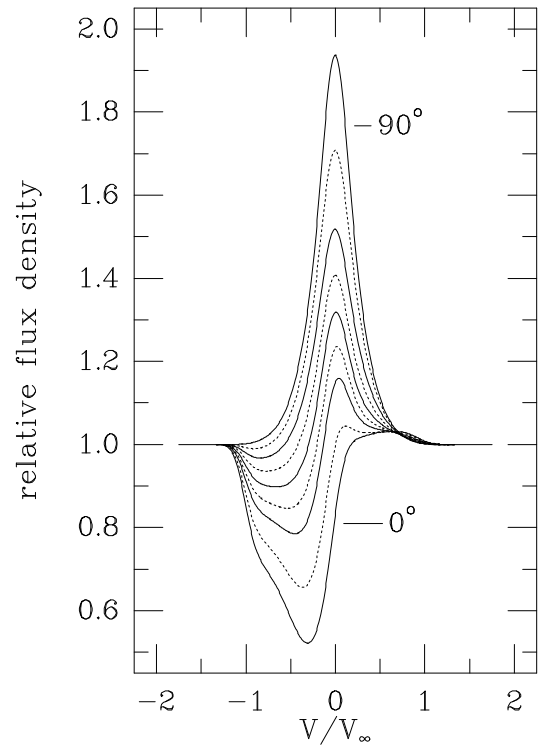
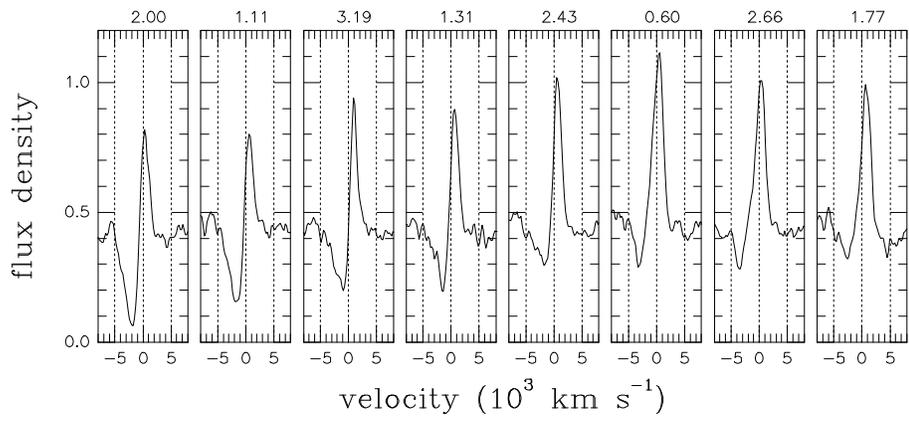
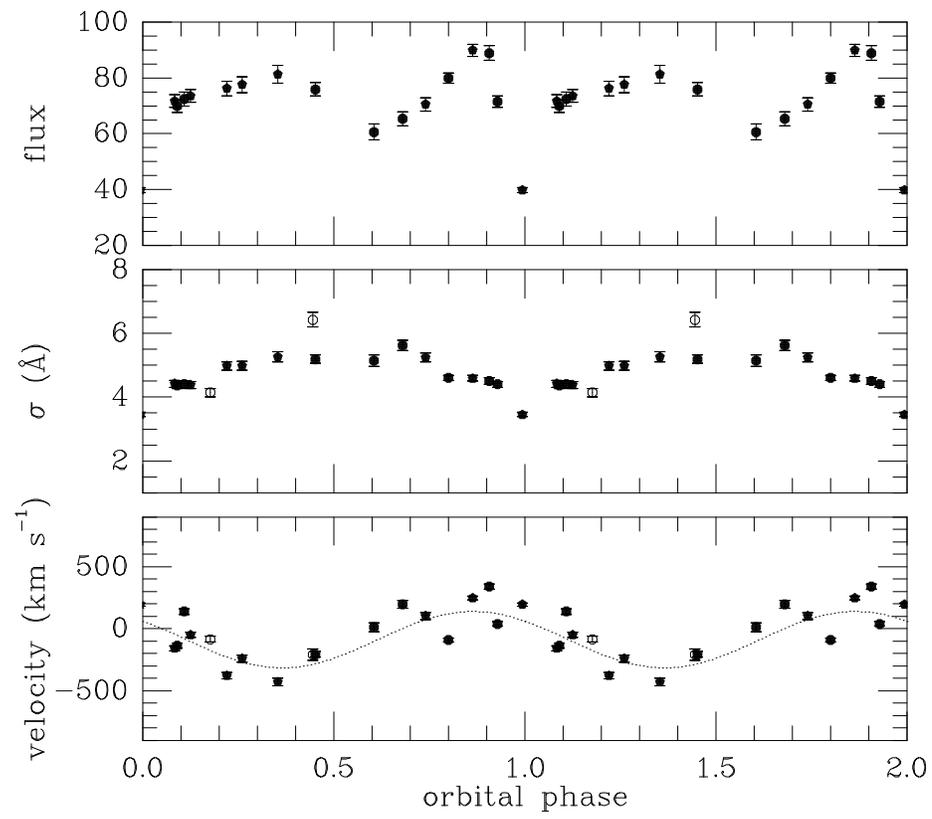
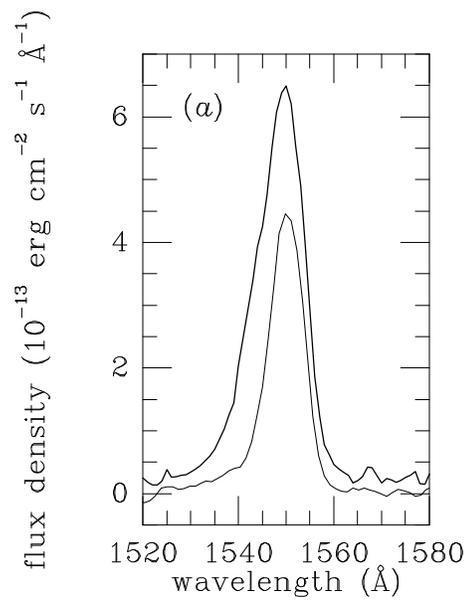
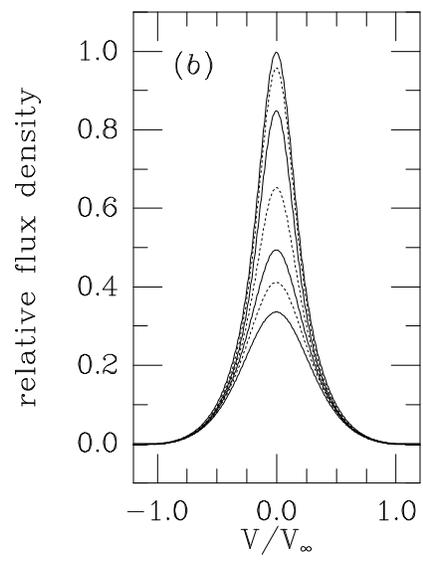
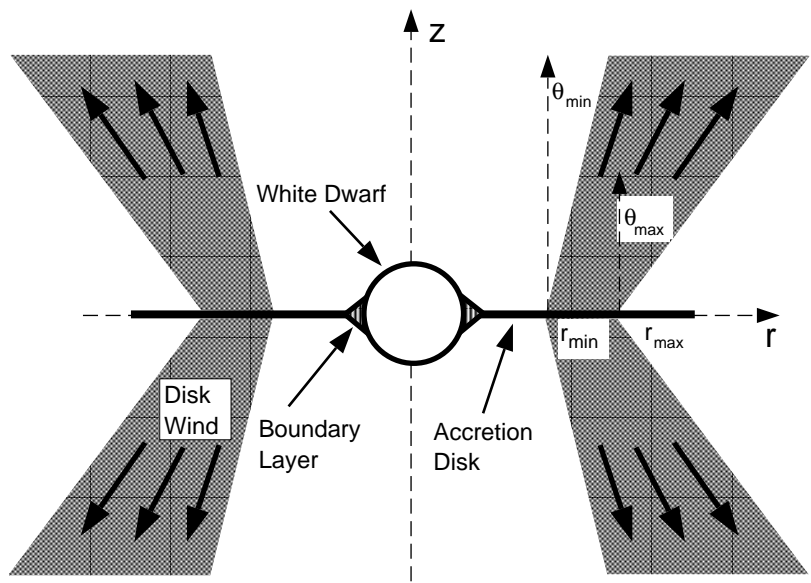
\bye